\documentclass[conference]{IEEEtran}
\IEEEoverridecommandlockouts
\usepackage{cite}
\usepackage{amsmath,amssymb,amsfonts}
\usepackage{algorithmic}
\usepackage{graphicx}
\usepackage{textcomp}

\usepackage{xcolor}
\def\BibTeX{{\rm B\kern-.05em{\sc i\kern-.025em b}\kern-.08em
    T\kern-.1667em\lower.7ex\hbox{E}\kern-.125emX}}
\begin{document}

\title{Singular Value Decomposition and Entropy Dimension of Fractals}

\author{\IEEEauthorblockN{Xiaojing Weng\IEEEauthorrefmark{1},
Altai Perry\IEEEauthorrefmark{2},
Michael Maroun\IEEEauthorrefmark{3}, and
Luat T. Vuong\IEEEauthorrefmark{1}\IEEEauthorrefmark{2}}
\IEEEauthorblockA{\IEEEauthorrefmark{1} Dept. Electrical and Computer Engineering, University of California, Riverside, CA USA}
\IEEEauthorblockA{\IEEEauthorrefmark{2} Dept. Mechanical Engineering, University of California, Riverside, CA USA}
\IEEEauthorblockA{\IEEEauthorrefmark{3} TeXDyn Industries Corporate Laboratories, Austin, TX USA}

\\
LuatV@UCR.edu
}

\maketitle

\begin{abstract}
We analyze the singular value decomposition (SVD) and SVD entropy of Cantor fractals produced by the Kronecker product. Our primary results show that SVD entropy is a measure of image ``complexity dimension" that is invariant under the number of Kronecker-product self-iterations {\it i.e., fractal order}. SVD entropy is therefore similar to the fractal Hausdorff complexity dimension but suitable for characterizing fractal wave phenomena. Our field-based normalization (Renyi entropy index = 1) illustrates the uncommon step-shaped and cluster-patterned distributions of the fractal singular values and their SVD entropy. As a modal measure of complexity, SVD entropy has uses for a variety of wireless communication, free-space optical, and remote sensing applications. 

\end{abstract}


\section{Introduction}
Fractals manifest in systems where the constituent interactions exhibit self-similar spatial and/or temporal growth dynamics \cite{mandelbrot1987fractals}. Spatiotemporal dynamics also evolve from the waves emanating (reflected, scattered, transmitted) from fractal apertures or envelopes undergoing diffraction. Such waves are referred to as `diffractals', a term coined by M.V. Berry to highlight distinct, non-Gaussian diffraction behavior \cite{Berry1979}. Fractal geometries and diffractals have attracted widespread attention in many branches of science with applications in engineering such as digital image processing (especially image compression \cite{jacquin1993fractal, zhao2005fractal} and antenna design \cite{Sundaram2007, puente1998behavior, sharma2017journey, werner2003overview, maraghechi2010enhanced}). We have recently studied fractal-encoded space-division multiplexing \cite{Weng2021, Moocarme2015}. For communication systems, the advantages of fractal antennas lie in broadband applications \cite{Wqrner2003, Manimegalai2009}. These advantages stem from the broadband electromagnetic response associated with multiple scales of spatial dimensions that are present within a fractal pattern. Most fractal applications exploit fractals' high level of information redundancy, which is organized in strongly-corrugated spatial patterns \cite{Korolenko2021, Verma2012, Verma2013}. 


Here, we focus on Cantor fractals that are generated from binary-valued kernels using the Kronecker product:
\begin{equation}
\mathbf{K}^{(n)} = \mathbf{K}^{(1)} \otimes  \mathbf{K}^{(1)}  \otimes \mathbf{K}^{(1)}   ... \,\,\,\text{($n$ times)}
\end{equation} 
where $n$ is the number of iterations of the Kronecker product, or the fractal order (FO), and $\mathbf{K}^{(1)}$ is a binary fractal kernel carrying `0's or `1's --- a square array of size $s \times s$. The fractal $\mathbf{K}^{(n)}$ with FO = $n$ has size $s^n \times s^n$. Our study of binary patterns is an important starting point for understanding more complex patterns. In fractal antenna applications, Cantor fractals are common since their binary structures are straightforward to fabricate. Additionally, for free space communication systems, Cantor fractals are suited to established intensity modulation/direct detection (IM/DD) schemes \cite{ghassemlooy2019optical, Moocarme2015, Weng2021}.  
\begin{figure}[bth]
\centerline{\includegraphics[width =\linewidth]{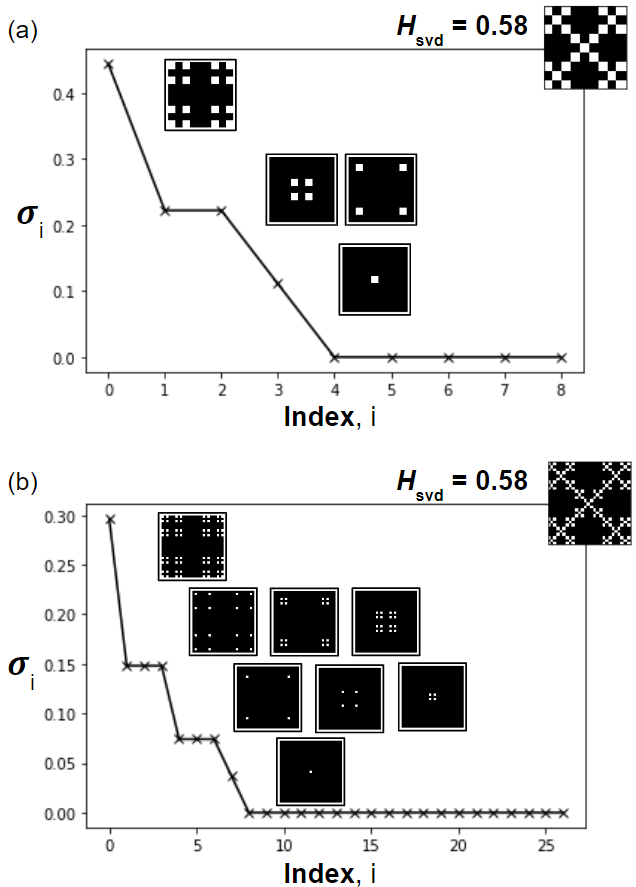}}
\caption{Singular values distributed from largest to smallest of an 3x3 `X'-kerneled fractal with fractal order (a) $n=2$ and (b) $n=3$. The insets are eigenimages ($\mathbf{S}_i$) corresponding to each singular value. The singular values of a fractal are combinatorial products of the singular values of its fundamental kernels. All Cantor fractals exhibit step-shaped singular value distributions. This `X'-kerneled fractal is a special case where the eigenimages are binary, but most have multi-valued eigenimages.}
\label{fig1test}
\end{figure}

We study the singular value decomposition (SVD) and SVD entropy of binary Cantor fractals generated via Kronecker product. A Cantor fractal provides an especially visual example that is represented with a few orthogonal modes, which can delineate families of Cantor fractals. The families are characterized by $n$ and $s$. The singular values are also unchanged with linear operations; Matrix $\mathbf{K}$ and its transform under a linear operator $\mathcal{F}$ are decomposed as:
\begin{eqnarray}
    \mathbf{K}&=&\mathbf{U\Sigma V} \label{svdo} \\
    \mathbf{\mathcal{F}K}&=&\mathcal{F}\mathbf{U\Sigma V} \label{svdf} \end{eqnarray}
where $\mathbf{\Sigma}$ is the same diagonal matrix in Eqs. \ref{svdo} and \ref{svdf} with positive singular values $\sigma_i$. While $\sigma_i$ remains invariant under $\mathbf{\mathcal{F}}$, the effective left $\mathbf{U}$ and right $\mathbf{V}$ similarity eigenmatrices may change and influence the eigenimages $\mathbf{S}_i$:
\begin{equation}
    (\mathbf{S}_i)_{jk} = {\mathbf{U}_{ji}\mathbf{V}_{ik}}.
\end{equation}
The primary purpose of this paper is to show that SVD entropy (${H}_{{\rm svd}}$) is a novel measure for Cantor fractals: 
\begin{eqnarray}
    {H}_{{\rm svd}} (\mathbf{K}^{(n)}) &=& \frac{1}{n\log_2 s}\sum_{i}^{s^n}p_i \log_2 p_i, \label{hsvdeq}
\end{eqnarray}
where we emphasize a specific probability  normalization ($\sum_i^{s^n} p_i = 1$) where $i = 1,\cdots,s^n$ and $p_i$ is related to the singular values:
\begin{equation}
    p_i = \frac{\sigma_i}{\sum_j^{s^n} \sigma_j}. \label{fnorm}
\end{equation}
Like the Hausdorff dimension and other entropy measures, SVD entropy is invariant under fractal order and therefore is useful for characterizing the fractal-dimension complexity. Moreover, since SVD entropy is also invariant under linear diffraction, it may be an appropriate measure for studies of diffractals. In Eq. \ref{hsvdeq}, we define an SVD-entropy measure that is similar to other entropy measures \cite{Alter2000}; it differs critically due to the normalization defined in Eq. \ref{fnorm}. Our normalization implies that we are treating each eigenimage as a modal transmission channel in the propagation of electromagnetic waves \cite{Miller2019, Choi2011}. 

Figure \ref{fig1test} shows the singular values and the modes associated with a fractal kernel $\mathbf{K}^{(1)} = \begin{bmatrix} 1 & 0& 1\\ 0 & 1 & 0\\1 & 0 & 1\\
\end{bmatrix}$. This particular symmetric kernel has eigenimages that are also binary-valued and it is unique in doing so: asymmetric fractal images results in multi-valued (non-binary) eigenimages. More generally, the step-valued singular values yield equivalent $H_{{\rm svd}}$ that does not depend on $n$. The proof of this invariance over fractal order is provided in Sec. \ref{Sec:BG}. We discuss the implications for diffractals and wire communication systems in Sec. \ref{discu}.

To our knowledge, little to no work has analyzed the singular measures of Cantor fractals.  For example, there is no rigorous characterization theorem that gives an all-encompassing presentation for the mathematical object as an integral of a continuous parameter like fractal order or kernel size. Here, we aim to connect digital image processing with wave fractal patterns for applications in wireless communications. 

\section{Background}
  \begin{figure}[h!]
\centerline{\includegraphics[width = \linewidth]{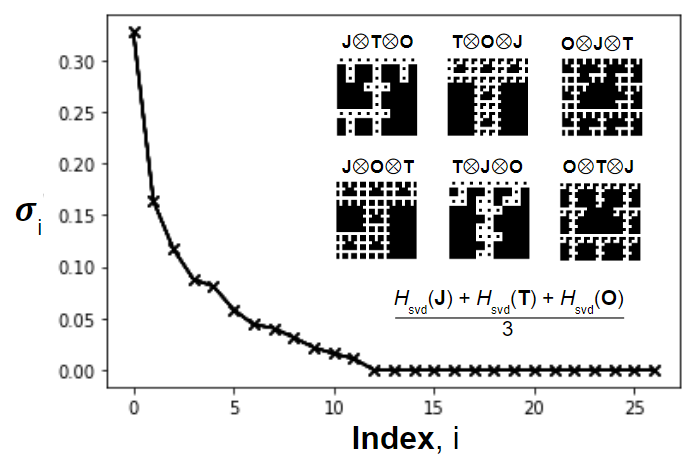}}
\caption{The images (inset) depend on the Kronecker-product order, but the distribution of singular values does not depend on product order. The $H_{{\rm svd}}$ is the average of that for constituent matrices.}
\label{fig2test}
 \end{figure}
Fractals are characterized by a complexity dimension, which describes how the fractal scales as the fractal grows (for example via an iterated function): both the box counting and Hausdorff dimensions are measures of how the surface area or volume scales (assuming a power law); other fractal dimensions (e.g., information or correlation dimensions) provide statistical measures of the spatial homogeneity and heterogeneity \cite{Young1982}; finally, entropy measures take into account the probabilities associated with the fractal points and vectors that map the fractal self-similar growth. 

Most entropy measures stem from a Boltzmann-Planck definition based on the system microstates and their probabilities. This definition is subsumed in a Renyi's generalized entropy measure \cite{Rnyi1959}:
\begin{equation}
    {H}_{\text{Renyi}} = \frac{1}{1-\alpha}\log \left( \sum_{i=1}^N{p_{i}^{\alpha}} \right) 
\end{equation}
where $\alpha$ is the entropy index. The limit as $\alpha \rightarrow 1$ identifies Shannon entropy \cite{rioul2021}, whereas the $\alpha=2$, referred to as collision entropy, provides an uncertainty relation for quantum observables \cite{Bosyk2012}. When $\alpha$ is large, entropy is dictated by states with higher probabilities. The index $\alpha$ is important for explaining the fractal dynamics.

The measures of fractal complexity based on local, spatial measures are in general not invariant under diffraction and diffusion. As a result, the fractal complexity dimension---while useful for understanding growth dynamics of the fractal itself---evolves as a function of propagation. Since waves diffract most strongly at edges, not volumes, space-domain measures associated with volume, homogeneity, and heterogeneity may not be useful for defining diffraction-related fractal wave phenomena, especially since the spatial profiles of diffractals evolve dramatically as they propagate to the far field \cite{Berry1979, Weng2021}. In contrast, a measure of fractal complexity based on modal measures is invariant under diffraction and diffusion.

Our definition of SVD entropy [Eqs. \ref{hsvdeq} and \ref{fnorm}] solves this issue by measuring both the self-similarity of fractals and their wave propagation. Our $H_{{\rm svd}}$ measure is suitable for characterizing the linear evolution and propagation of diffractals since $H_{{\rm svd}}$ is invariant under linear operations. This approach diverges from the conventional definition of SVD entropy \cite{Alter2000}, which uses an entropy index of $\alpha = 2$ so that the probabilities are $p_i = \frac{\sigma_i^2}{\sum_i^{s^n} \sigma_i^2}$. Additionally, most SVD-based probabilities for optics are also based on intensity sensor measurements, which employ a multimodal entropy index of $\alpha = 2$ \cite{Miller2019}. By contrast, we study the $H_{{\rm svd}}$ with $\alpha = 1$ [Eq. \ref{fnorm}], which provides a measure of the electric fields. This $H_{{\rm svd}}$ is also compatible with a transmission-matrix description of wave propagation through random media; the distribution of singular values describe  the distribution of open transmission channels \cite{Choi2011} and a larger $H_{{\rm svd}}$ implies a wider distribution of open transmission channels.

\section{results}

\subsection{Fractal step distributions of singular values}
Consider the singular values of the Kronecker product \cite{Schacke04onthe, Brewer1978}:
\begin{eqnarray}
\mathbf{C} &=& \mathbf{A} \otimes \mathbf{B}\\
&=&  (\mathbf{U}^{(A)}\otimes \mathbf{U}^{(B)})(\mathbf{\Sigma}^{(A)} \otimes \mathbf{\Sigma}^{(B)})(\mathbf{V}^{(A)}\otimes \mathbf{V}^{(B)})\nonumber \\
& =& \mathbf{U}^{(C)} \mathbf{\Sigma}^{(C)}\mathbf{V}^{(C)}.
\end{eqnarray}

By inspection, the singular value matrices are related: $\Sigma^{(C)} = \mathbf{\Sigma}^{(A)} \otimes \mathbf{\Sigma}^{(B)}$. The singular values $\sigma_k^{(C)}$ are the diagonal entries of $\mathbf{\Sigma}^{(C)}$ \cite{Schacke04onthe}. We identify the eigenvalues of the product in terms of those of the constituent matrices:

\begin{equation}
    \sigma^{(C)}_{k} = \sigma^{(A)}_i \sigma^{(B)}_j \;\; \label{svi}
\end{equation}

where $i=1,\cdots,N_A, j=1,\cdots,N_B$, and $k = 1,\cdots,N_A N_B$. Here, $N_A$ and $N_B$ are the dimensions of $\mathbf{A}$ and $\mathbf{B}$. Each iterated product increases the number of spatial modes and number of singular values. 

When $\mathbf{A}=\mathbf{B}$, a step distribution of singular values is observed [Fig.\ref{fig1test}]. The duplicated singular values in each step arise because of the symmetry of the combination of the kernel singular values, which is $\sigma^{(A)}_i \sigma^{(A)}_j = \sigma^{(A)}_j \sigma^{(A)}_i$, when $i \ne j$. The singular value is unique when $i=j$. In Fig. \ref{fig1test}, since the kernel has 2 unique singular values, the count of the duplicated values are represented by a binomial coefficient. If the kernel has 3 unique, nonzero singular values, then the number of duplicates at each step value would be represented by trinomial coefficients \cite{Keeney1969}. In Sec. \ref{Sec:BG}, we derive that $H_{{\rm svd}}$ is invariant under fractal order.

\begin{figure}[hbt]
\centerline{\includegraphics[width = \linewidth]{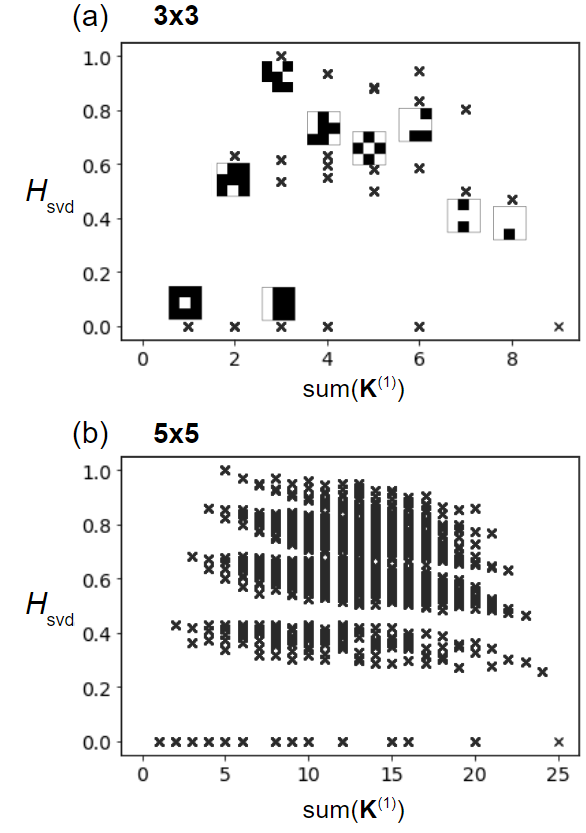}}
\caption{ Discrete values of the $H_{{\rm svd}}$ as a function of fill factor, or sum($\mathbf{K}^{(1)}$), for (a) 3x3 and (b) 5x5 kernels. For binary-valued fractals, the highest values of  $H_{{\rm svd}}$ have lower fill factor. }
\label{fig3}
\end{figure}
 
\begin{figure*}[hbt]
    \centering
    \includegraphics[width=\textwidth]{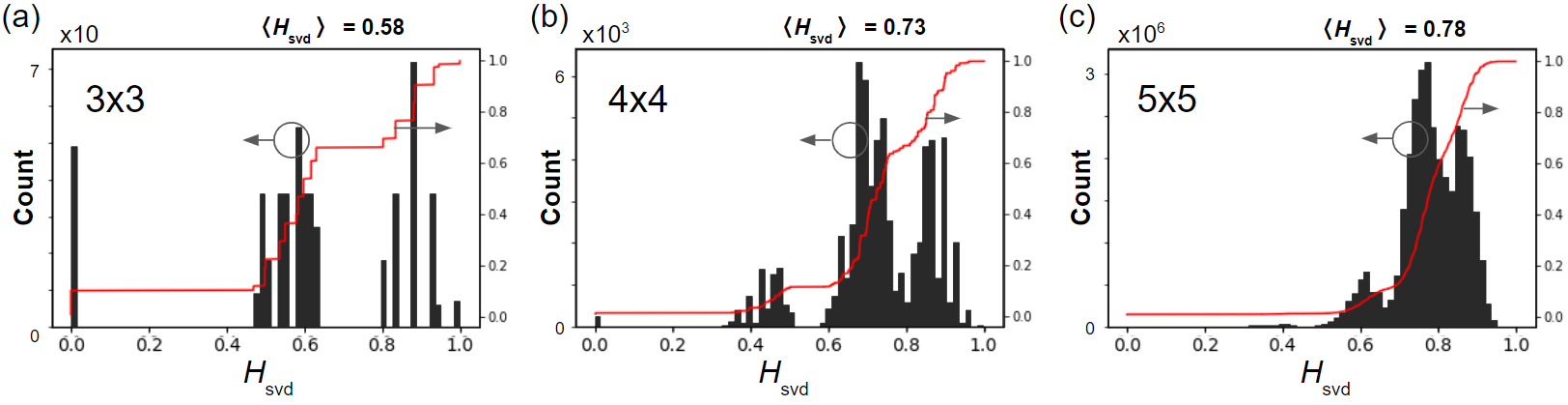}
\caption{Histogram of $H_{{\rm svd}}$ (black) and cumulative density function (red) showing clusters. The mean value of the $H_{{\rm svd}}$ increases with the size of the input kernel and the clusters become less distinct.  
}
\label{fig4}
\end{figure*}

\subsection{$H_{{\rm svd}}$: a constituent average, invariant under fractal order \label{Sec:BG}}

To show that $H_{{\rm svd}}$ is invariant under the iteration numbers of the Kronecker product, consider Eq. \ref{svi} and the normalization according to Eq. \ref{fnorm}:
\begin{eqnarray}
p^{(C)}_k &=& \frac{ {\sigma^{(A)}_{i}\sigma^{(B)}_{j}}}{{\sum_{r}^{N_A}\sum_{q}^{N_B} \sigma^{(A)}_{r}\sigma^{(B)}_{q}} 
}\;\; \text{where }k = ij\\
&=& p_i^{(A)}p_j^{(B)} 
\end{eqnarray}
 
Starting from Eq. \ref{p1} and ($\sum_i^N p_i = 1$) :
\begin{eqnarray}
    {H}_{{\rm svd}}(\mathbf{C})&=& \frac{\sum_k^{N_AN_B}p^{(C)}_k \log_2p^{(C)}_k }{\log_2 N_AN_B} \label{p1}\\
    &=& \frac{\sum_i^{N_A}\sum_j^{N_B}p^{(A)}_ip^{(B)}_j \log_2p^{(A)}_ip^{(B)}_j }{\log_2 N_AN_B} \nonumber\\
    &=& \frac{\sum_{j}^{N_B}  p_j^{(B)}\log_2p_j^{(B)} 
    + \sum_{i}^{N_A}  p_i^{(A)}\log_2p_i^{(A)} }{\log_2 N_AN_B} \nonumber\\     
    &=& \frac{\log_2 N_A {H}_{{\rm svd}}(\mathbf{A}) 
    + \log_2 N_B{H}_{{\rm svd}}(\mathbf{B})  }{\log_2 N_AN_B} \label{fif}
\end{eqnarray}
From Eq. \ref{fif}, if $\mathbf{A}$ and $\mathbf{B}$ are the same size, or $N_A = N_B$: 
\begin{equation}
    {H}_{{\rm svd}}(\mathbf{A} \otimes \mathbf{B}) =\frac{{H}_{{\rm svd}}(\mathbf{A})+ {H}_{{\rm svd}}(\mathbf{B}) }{2}
\end{equation}
\\
Moreover, if $\mathbf{A} = \mathbf{B} =\mathbf{K}^{(1)}$, then 
\begin{equation}
    {H}_{{\rm svd}}(\mathbf{K}^{(2)}) =  \frac{{H}_{{\rm svd}}(\mathbf{K}^{(1)})+{H}_{{\rm svd}}(\mathbf{K}^{(1)})}{2} = {H}_{{\rm svd}}(\mathbf{K}^{(1)}).
\end{equation} 
By extension, the $H_{{\rm svd}}$ of any Cantor fractal is invariant under fractal order $n$. 

The $H_{{\rm svd}}$ of an image produced by the Kronecker product does not depend on the Kronecker-product order (unlike the image itself). In fact, if the constituents have the same size, then $H_{{\rm svd}}$ is the average of the $H_{{\rm svd}}$ of the constituent matrices [Fig. \ref{fig2test}] and the singular values themselves  do not depend on the Kronecker-product order. 
Figure \ref{fig2test} shows the distribution of singular values of fractals from different constituent matrices and their Kronecker-product ordering. In contrast to Cantor fractals [Fig. \ref{fig1test}], these distributions are smooth since there are no longer unique pairings of the constituent singular values. In other words, the steps in Fig. \ref{fig1test} point to the redundancy of fractal patterns.

 \subsection{Higher $H_{{\rm svd}}$ with lower fill factor and with larger kernels}
 
$H_{{\rm svd}}$ values are discrete measures that group kernel patterns into families. Higher values of $H_{{\rm svd}}$ are generally calculated from matrices with a higher number of internal edges. This is why kernels with lower fill factor (as measured by sum($\mathbf{K}^{(1)}$)) have higher $H_{{\rm svd}}$. Figure \ref{fig3} shows the $H_{{\rm svd}}$ values for the 2$^9$, $s=3$ kernels and 2$^{25}$, $s=5$ kernels. There are less than 20 unique, or distinct values for the $H_{{\rm svd}}$ of 3x3 kernels, and less than 100 for 5x5 kernels. Including the $H_{{\rm svd}} = 0$, there are $s$ groups of $H_{{\rm svd}}$ clusters for $s \times s$ kernels.  

The groups of $H_{{\rm svd}}$ values are also visible in Fig. \ref{fig4}, which shows histograms and cumulative density distributions of $H_{{\rm svd}}$ for $s$ = 3, 4, and 5. In spite of the non-Gaussian patterns and strongly grouped clusters of $H_{{\rm svd}}$ values, the distribution median and mean are approximately the same.  As the kernel size $s$ increases, the mean $H_{{\rm svd}}$ increases; as the number of kernel pixels increases, there are a greater number of kernel combinations with higher $H_{{\rm svd}}$. Kernels with low fill factor (where sum$(\mathbf{K}^{(1)})$ $<s^2/2$) tend to have both the highest and lowest values of $H_{{\rm svd}}$.

\section{Discussion \label{discu}}
$H_{{\rm svd}}$ reveals the distribution of orthogonal spatial modes or eigenimages needed to represent matrix data. 

It is invariant under the Kronecker product and provides a novel measure of the information contained in matrix data (that is, the kernel data). However as a complexity dimension for fractals, there are differences to consider between $H_{{\rm svd}}$ and other fractal complexity measures. For example, higher-order Cantor fractals of the identity matrix produce a thin line ($H_{{\rm svd}}$=1), which we would not generally associate with fractal complexity by any existing measure. If we can reconcile this difference in measures, then $H_{{\rm svd}}$ may be useful for studying diffractals because $H_{{\rm svd}}$ is invariant under linear transformations like diffraction. On one hand, it is surprising that $H_{{\rm svd}}$ is invariant under both linear spatial transformations and fractal order, which implies nonlinear dynamics. On the other hand, we have limited our study to Cantor fractals defined by Kronecker product. Also, the invariance of $H_{{\rm svd}}$ is related to our choice of $\alpha = 1$ in our normalization Eq. \ref{fnorm}, which is not the convention set forth by earlier work \cite{Alter2000}.

Our normalization is a field-based measure rather than an intensity-based measure. Given that it is a challenge to measure the phase of optical signals, $H_{{\rm svd}}$ may be a difficult characteristic to measure in practice. At the same time, our knowledge of $H_{{\rm svd}}$ relates to our understanding of transmission matrices and the propagation of waves in random media, where $H_{{\rm svd}}$ quantifies the distribution of open transmission channels. Our work indicates that the distribution of open transmission channels for a Cantor fractal aperture increases with the kernel $H_{{\rm svd}}$.

\section{Conclusion}  
We calculate $H_{{\rm svd}}$ as an entropy measure with normalized singular values [$\alpha = 1$, Eq. \ref{fnorm}]. $H_{{\rm svd}}$ characterizes the original kernel data of Cantor fractals defined via iterations of the Kronecker product. As the fractal order increases, the number of singular values increases while step-shaped distributions of these singular values emerge, but the $H_{{\rm svd}}$ does not change. We show that the distributions of singular values for 2D Cantor fractals are step-shaped, where the number of duplicate singular values depend on the initial kernel's singular values. These duplicates point to the well-known but understudied redundant patterns of fractals, which may also be useful in further studies of diffractals.

\section*{acknowledgments}

Authors acknowledge funding from DARPA YFA D19AP00036. 

\bibliographystyle{IEEEtran}
\bibliography{IEEEabrv,SVDH}

\begin{thebibliography}{10}
\providecommand{\url}[1]{#1}
\csname url@samestyle\endcsname
\providecommand{\newblock}{\relax}
\providecommand{\bibinfo}[2]{#2}
\providecommand{\BIBentrySTDinterwordspacing}{\spaceskip=0pt\relax}
\providecommand{\BIBentryALTinterwordstretchfactor}{4}
\providecommand{\BIBentryALTinterwordspacing}{\spaceskip=\fontdimen2\font plus
\BIBentryALTinterwordstretchfactor\fontdimen3\font minus
  \fontdimen4\font\relax}
\providecommand{\BIBforeignlanguage}[2]{{%
\expandafter\ifx\csname l@#1\endcsname\relax
\typeout{** WARNING: IEEEtran.bst: No hyphenation pattern has been}%
\typeout{** loaded for the language `#1'. Using the pattern for}%
\typeout{** the default language instead.}%
\else
\language=\csname l@#1\endcsname
\fi
#2}}
\providecommand{\BIBdecl}{\relax}
\BIBdecl

\bibitem{mandelbrot1987fractals}
B.~B. Mandelbrot and M.~Frame, ``Fractals,'' \emph{Encyclopedia of physical
  science and technology}, vol.~5, pp. 579--593, 1987.

\bibitem{Berry1979}
\BIBentryALTinterwordspacing
M.~V. Berry, ``Diffractals,'' \emph{Journal of Physics A: Mathematical and
  General}, vol.~12, no.~6, pp. 781--797, Jun. 1979. [Online]. Available:
  \url{https://doi.org/10.1088/0305-4470/12/6/008}
\BIBentrySTDinterwordspacing

\bibitem{jacquin1993fractal}
A.~E. Jacquin, ``Fractal image coding: A review,'' \emph{Proceedings of the
  IEEE}, vol.~81, no.~10, pp. 1451--1465, 1993.

\bibitem{zhao2005fractal}
E.~Zhao and D.~Liu, ``Fractal image compression methods: A review,'' in
  \emph{Third International Conference on Information Technology and
  Applications (ICITA'05)}, vol.~1.\hskip 1em plus 0.5em minus 0.4em\relax
  IEEE, 2005, pp. 756--759.

\bibitem{Sundaram2007}
\BIBentryALTinterwordspacing
A.~Sundaram, M.~Maddela, and R.~Ramadoss, ``Koch-fractal folded-slot antenna
  characteristics,'' \emph{{IEEE} Antennas and Wireless Propagation Letters},
  vol.~6, pp. 219--222, 2007. [Online]. Available:
  \url{https://doi.org/10.1109/lawp.2007.895293}
\BIBentrySTDinterwordspacing

\bibitem{puente1998behavior}
C.~Puente-Baliarda, J.~Romeu, R.~Pous, and A.~Cardama, ``On the behavior of the
  sierpinski multiband fractal antenna,'' \emph{IEEE Transactions on Antennas
  and propagation}, vol.~46, no.~4, pp. 517--524, 1998.

\bibitem{sharma2017journey}
N.~Sharma and V.~Sharma, ``A journey of antenna from dipole to fractal: A
  review,'' \emph{Journal of Engineering Technology}, vol.~6, no.~2, pp.
  317--351, 2017.

\bibitem{werner2003overview}
D.~H. Werner and S.~Ganguly, ``An overview of fractal antenna engineering
  research,'' \emph{IEEE Antennas and propagation Magazine}, vol.~45, no.~1,
  pp. 38--57, 2003.

\bibitem{maraghechi2010enhanced}
P.~Maraghechi and A.~Elezzabi, ``Enhanced thz radiation emission from plasmonic
  complementary sierpinski fractal emitters,'' \emph{Optics express}, vol.~18,
  no.~26, pp. 27\,336--27\,345, 2010.

\bibitem{Weng2021}
\BIBentryALTinterwordspacing
X.~Weng and L.~T. Vuong, ``Fractal, diffraction-encoded space-division
  multiplexing for {FSO} with misalignment-robust, roaming transceivers,'' Dec.
  2021. [Online]. Available: \url{https://doi.org/10.21203/rs.3.rs-1146919/v1}
\BIBentrySTDinterwordspacing

\bibitem{Moocarme2015}
\BIBentryALTinterwordspacing
M.~Moocarme and L.~T. Vuong, ``Robustness and spatial multiplexing via
  diffractal architectures,'' \emph{Optics Express}, vol.~23, no.~22, p. 28471,
  Oct. 2015. [Online]. Available: \url{https://doi.org/10.1364/oe.23.028471}
\BIBentrySTDinterwordspacing

\bibitem{Wqrner2003}
\BIBentryALTinterwordspacing
D.~Wqrner and S.~Ganguly, ``An overview of fractal antenna engineering
  research,'' \emph{{IEEE} Antennas and Propagation Magazine}, vol.~45, no.~1,
  pp. 38--57, Feb. 2003. [Online]. Available:
  \url{https://doi.org/10.1109/map.2003.1189650}
\BIBentrySTDinterwordspacing

\bibitem{Manimegalai2009}
\BIBentryALTinterwordspacing
B.~Manimegalai, S.~Raju, and V.~Abhaikumar, ``A multifractal cantor antenna for
  multiband wireless applications,'' \emph{{IEEE} Antennas and Wireless
  Propagation Letters}, vol.~8, pp. 359--362, 2009. [Online]. Available:
  \url{https://doi.org/10.1109/lawp.2008.2000828}
\BIBentrySTDinterwordspacing

\bibitem{Korolenko2021}
\BIBentryALTinterwordspacing
P.~V. Korolenko, R.~T. Kubanov, and A.~Y. Mishin, ``Features of the complex
  representation of diffractal wave structures,'' \emph{Bulletin of the Russian
  Academy of Sciences: Physics}, vol.~85, no.~1, pp. 53--56, Jan. 2021.
  [Online]. Available: \url{https://doi.org/10.3103/s1062873821010160}
\BIBentrySTDinterwordspacing

\bibitem{Verma2012}
\BIBentryALTinterwordspacing
R.~Verma, V.~Banerjee, and P.~Senthilkumaran, ``Redundancy in cantor
  diffractals,'' \emph{Optics Express}, vol.~20, no.~8, p. 8250, Mar. 2012.
  [Online]. Available: \url{https://doi.org/10.1364/oe.20.008250}
\BIBentrySTDinterwordspacing

\bibitem{Verma2013}
\BIBentryALTinterwordspacing
R.~Verma, M.~K. Sharma, V.~Banerjee, and P.~Senthilkumaran, ``Robustness of
  cantor diffractals,'' \emph{Optics Express}, vol.~21, no.~7, p. 7951, Mar.
  2013. [Online]. Available: \url{https://doi.org/10.1364/oe.21.007951}
\BIBentrySTDinterwordspacing

\bibitem{ghassemlooy2019optical}
Z.~Ghassemlooy, W.~Popoola, and S.~Rajbhandari, \emph{Optical wireless
  communications: system and channel modelling with
  Matlab{\textregistered}}.\hskip 1em plus 0.5em minus 0.4em\relax CRC press,
  2019.

\bibitem{Alter2000}
\BIBentryALTinterwordspacing
O.~Alter, P.~O. Brown, and D.~Botstein, ``Singular value decomposition for
  genome-wide expression data processing and modeling,'' \emph{Proceedings of
  the National Academy of Sciences}, vol.~97, no.~18, pp. 10\,101--10\,106,
  Aug. 2000. [Online]. Available:
  \url{https://doi.org/10.1073/pnas.97.18.10101}
\BIBentrySTDinterwordspacing

\bibitem{Miller2019}
\BIBentryALTinterwordspacing
D.~A.~B. Miller, ``Waves, modes, communications, and optics: a tutorial,''
  \emph{Adv. Opt. Photon.}, vol.~11, no.~3, pp. 679--825, Sep 2019. [Online].
  Available:
  \url{http://www.osapublishing.org/aop/abstract.cfm?URI=aop-11-3-679}
\BIBentrySTDinterwordspacing

\bibitem{Choi2011}
\BIBentryALTinterwordspacing
W.~Choi, A.~P. Mosk, Q.-H. Park, and W.~Choi, ``Transmission eigenchannels in a
  disordered medium,'' \emph{Physical Review B}, vol.~83, no.~13, Apr. 2011.
  [Online]. Available: \url{https://doi.org/10.1103/physrevb.83.134207}
\BIBentrySTDinterwordspacing

\bibitem{Young1982}
\BIBentryALTinterwordspacing
L.-S. Young, ``Dimension, entropy and lyapunov exponents,'' \emph{Ergodic
  Theory and Dynamical Systems}, vol.~2, no.~1, pp. 109--124, Mar. 1982.
  [Online]. Available: \url{https://doi.org/10.1017/s0143385700009615}
\BIBentrySTDinterwordspacing

\bibitem{Rnyi1959}
\BIBentryALTinterwordspacing
A.~R{\'{e}}nyi, ``On the dimension and entropy of probability distributions,''
  \emph{Acta Mathematica Academiae Scientiarum Hungaricae}, vol.~10, no. 1-2,
  pp. 193--215, Mar. 1959. [Online]. Available:
  \url{https://doi.org/10.1007/bf02063299}
\BIBentrySTDinterwordspacing

\bibitem{rioul2021}
\BIBentryALTinterwordspacing
O.~Rioul, ``{This is IT: A Primer on Shannon's Entropy and Information},'' in
  \emph{{Information Theory Poincar{\'e} Seminar 2018}}, ser. Progress in
  Mathematical Physics, B.~Duplantier and V.~Rivasseau, Eds.\hskip 1em plus
  0.5em minus 0.4em\relax {Birkh{\"a}user, Springer Nature}, Aug. 2021,
  vol.~78. [Online]. Available: \url{https://hal.telecom-paris.fr/hal-03326385}
\BIBentrySTDinterwordspacing

\bibitem{Bosyk2012}
\BIBentryALTinterwordspacing
G.~M. Bosyk, M.~Portesi, and A.~Plastino, ``Collision entropy and optimal
  uncertainty,'' \emph{Physical Review A}, vol.~85, no.~1, Jan. 2012. [Online].
  Available: \url{https://doi.org/10.1103/physreva.85.012108}
\BIBentrySTDinterwordspacing

\bibitem{Schacke04onthe}
K.~Schäcke, ``On the kronecker product,'' 2004.

\bibitem{Brewer1978}
\BIBentryALTinterwordspacing
J.~Brewer, ``Kronecker products and matrix calculus in system theory,''
  \emph{{IEEE} Transactions on Circuits and Systems}, vol.~25, no.~9, pp.
  772--781, Sep. 1978. [Online]. Available:
  \url{https://doi.org/10.1109/tcs.1978.1084534}
\BIBentrySTDinterwordspacing

\bibitem{Keeney1969}
\BIBentryALTinterwordspacing
R.~L. Keeney and J.~F. Ramaley, ``On the trinomial coefficients,''
  \emph{Mathematics Magazine}, vol.~42, no.~4, pp. 210--213, Sep. 1969.
  [Online]. Available: \url{https://doi.org/10.1080/0025570x.1969.11975963}
\BIBentrySTDinterwordspacing

\end{thebibliography}
\vspace{12pt}

\end{document}